\documentclass{article}
\usepackage{spconf,amsmath,graphicx}
\usepackage{bm}
\usepackage{setspace}
\usepackage{subfigure}
\usepackage{here}
\usepackage{amssymb}

\title{W-NET BF: DNN-BASED BEAMFORMER USING JOINT TRAINING APPROACH}
\name{Yuichiro Koyama$^{\dagger \star}$, Bhiksha Raj$^\dagger$}
\address{$^\dagger$ Carnegie Mellon University, Pittsburgh, PA, USA\\ $^\star$ Sony Corporation, Minato-ku, Tokyo, Japan}

\begin{document}
\ninept
\maketitle

\begin{spacing}{0.89}
\begin{abstract}
Acoustic beamformers have been widely used to enhance audio signals. The best current methods are DNN-powered variants of the {\em generalized eigenvalue} (GEV) beamformer, and DNN-based {\em filter-estimation} methods that directly compute beamforming filters. Both approaches, while effective, have blindspots in their generalizability. We propose a novel approach that combines both approaches into a single framework that attempts to exploit the best features of both. The resulting model, called a W-Net beamformer, includes two components: the first computes a noise-masked reference which the second uses to estimate beamforming filters. Results on data that include a wide variety of room and noise conditions, including static and mobile noise sources, show that the proposed beamformer outperforms other methods in all tested evaluation metrics.

\end{abstract}


\section{Introduction}
\label{sec:intro}
As an effective enhancer of audio signals, the acoustic beamformer, which combines signals captured by an array of microphones to produce the enhanced signal, is widely used for various applications such as 
automatic speech recognition (ASR), speaker recognition, and hearing aids
\cite{barker2015third,vincent2017analysis,barker2018fifth,movsner2018dereverberation,jensen2015analysis}. The quest for improved beamformer algorithms remains an active area of research. 

Older, ``traditional'' beamformer algorithms required accurate preliminary detection of the direction of target sources \cite{griffiths1982alternative, capon1969high, bitzer2001superdirective},
a task fraught with peril. More recent methods have employed the GEV beamformer \cite{warsitz2007blind}, which does not require explicit source localization, but instead directly ``beamforms'' to directions of highest signal to noise ratio (SNR). The need for explicit localization is now, however, replaced by the equally challenging task of computing accurate cross-power spectral density matrices for noise.

In following the trend from many fields of AI, this problem has been found to be amenable to solutions based on deep learning. The empirically most successful methods have been based on the {\em mask estimation} approach, which estimates a {\em time-frequency mask} that localizes spectro-temporal regions of the signal that are dominated by noise, to compute its cross-power spectral density \cite{heymann2016neural, zhou2018robust, higuchi2018frame, wang2018mask, liu2018neural}. However, the approach comes with its drawbacks. The identification of ``noise'' and ``speech'' time-frequency components is based, in part, on heuristic thresholds; the estimators are trained based on these ad-hoc thresholds. 
Additionally, the computation of the spectral density matrices combines information from the entire recording, effectively assuming that the noise and source are spatially stationary, and tracking mobile sources remains a challenge \cite{Bddeker2018ExploringPA}.

An alternate DNN-based approach, the {\em filter-estimation} approach, directly estimates filters in a ``filter-and-sum'' beamformer \cite{johnson1993array} using a DNN. This approach attempts to handle spatio-temporal non-stationarity of signals through recurrent architectures that update filter parameters with time \cite{xiao2016study, meng2017deep, pfeifenberger2019deep}. While this is very effective when test conditions are similar to those seen in training data, it lacks the greater robustness and generalization of the mask-based method, where the DNN only performs the much simpler task of mask estimation and thus generalizes better.

In this paper we suggest a ``best-of-both-worlds'' beamforming solution, that combines the most useful aspects of both of the above approaches. We propose a two-stage model, where the first stage computes the equivalent of a masked spectrogram, which is then used as a reference by the second stage which computes the actual beamforming filters. It combines the superior generalization of the mask-based approach with the more optimal processing of the filter approach. Experiments show that the proposed method outperforms either approach on both static and mobile sound sources.

\end{spacing}
\begin{spacing}{0.89}

\section{Background and related work}
\label{sec:format}

The beamforming solution operates on signals captured by an {\em array} of microphones. Each microphones receives a slightly different version of the signal, influenced by a different room impulse response, and by different noise. Consider an array with $M$ microphones. The signal captured by the $m^{\rm th}$ microphone in the array can be written in the time-frequency domain as
\begin{equation}
X_m(t,f) = H_m(f)S(t,f) + N_m(t,f),
\end{equation}
where $S(t,f)$ represents the complex time-frequency coefficients of the clean speech, $N_m(t,f)$ represent the noise at the microphone, $H_m(f)$ is the (possibly time-varying) room impulse response at the microphone, and $X_m(t,f)$ is the (time-frequency characterization of) noisy signal actually captured by the microphone. The task of the beamformer is to combine the signals $X_m(t,f)$ in a manner such that the combined signal is as close to $S(t,f)$ as possible.

The most successful approach to beamforming is the {\em filter-and-sum} approach, in which each of the microphone signals is processed by a separate filter, and the filtered signals are subsequently summed to obtain the enhanced signal:
\begin{equation}
\hat{S}(t,f) = \bm{w}^H(f) \bm{x}(t,f).
\label{eq:FandS}
\end{equation}
In the equation above $\bm{w}(f) = [W_1(f), \cdots, W_M(f)]^\top$, where $W_m(f)$ represents the filter applied to the $m^{\rm th}$ array signal, and $\bm{x}(t,f) = [X_1(t,f), \cdots, X_M(t,f)]^\top$. The challenge is to estimate the filter $\bm{w}(f)$ that result in the $\hat{S}(t,f)$ with the highest SNR.

\subsection{Traditional Solutions}
The ``traditional'' solutions to beamforming estimate $\bm{w}(f)$ to optimize various objective measures computed from the known location of the source (the ``look'' direction), and various criteria related to the SNR of the enhanced signal, such as the SNR under spatially uncorrelated noise \cite{johnson1993array}, the ratio of the energy from the look direction to that from other directions \cite{bitzer2001superdirective}, minimum noise variance \cite{capon1969high}, and minimization of signal energy from non-look directions \cite{griffiths1982alternative,frost1972algorithm}. These methods generally require knowledge of the source direction and are sensitive to errors in estimating it.

\subsection{GEV and mask based beamformers}
The GEV beamformer \cite{warsitz2007blind} bypasses the requirement of knowledge of the look direction by estimating the filter parameters to maximize the signal to noise ratio:
\begin{equation}
\text{SNR}(f) = \frac{\bm{w}^H(f) \bm{\Phi}_S(f) \bm{w}(f) }{\bm{w}^H(f) \bm{\Phi}_N(f) \bm{w}(f)}.
\end{equation}
where $\bm{\Phi}_S(f)$ and $\bm{\Phi}_N(f)$ are the {\em cross-power} spectral matrices for speech and noise respectively. Maximization of the above objective results in a generalized eigenvalue solution (whence the name of the approach). The solution requires estimation of $\bm{\Phi}_S(f)$ and $\bm{\Phi}_N(f)$, which requires prior knowledge of speech- and noise-dominated time-frequency regions of the array signals.

The {\em mask-based} method \cite{heymann2016neural, zhou2018robust, higuchi2018frame, wang2018mask, liu2018neural} resolve this problem by estimating {\em masks} $r_N(t,f)$ and $r_S(t,f)$, computed using a DNN, which assign to any time-frequency location $(t,f)$ the probability of being noise and speech dominated respectively. Subsequently the cross-power spectral matrices are computed as
\begin{equation}
\bm{\Phi}_S(f) = \frac{1}{T}\sum_{t=1}^{T}[r_S(t,f)\bm{x}(t,f)\bm{x}^H(t,f)],
\end{equation}
\begin{equation}
\bm{\Phi}_N(f) = \frac{1}{T}\sum_{t=1}^{T}[r_N(t,f)\bm{x}(t,f)\bm{x}^H(t,f)].
\end{equation}
The DNNs that compute $r_S(t,f)$ and $r_N(t,f)$ are trained from signals with known noise. A heuristically chosen threshold $\theta$ is applied to each time-frequency component of the training signals, to decide whether to label it as noise or signal dominated, to train the DNN. 

\subsection{Beamformers with directly estimated filters}
The filter-estimation approach directly applies a DNN, typically a recurrent neural network employing LSTMs, to the multi-channel inputs to compute time-varying beamforming filters $\bm{w}(t,f)$ \cite{xiao2016study, meng2017deep, pfeifenberger2019deep}. The network is trained on multi-channel signals, with targets provided either by a simultaneously recorded clean signal or by higher-level models such as an ASR system \cite{meng2017deep}.
\\

We have already explained the relative merits and demerits of the methods: traditional methods need knowledge of the look direction, GEV methods estimate a single $\bm{w}(f)$ for the entire signal, and while filter-based methods do estimate time-varying filters, they are biased towards noises and environments seen in training, since no additional cues relating to the noise in the {\em current} test condition, besides the noisy signal itself, are provided to the algorithms.

The method we propose below addresses these issues.

\section{Proposed Method}
\label{sec:method}
\begin{figure}[htb]
  \centering
  \centerline{\includegraphics[width=8.0cm]{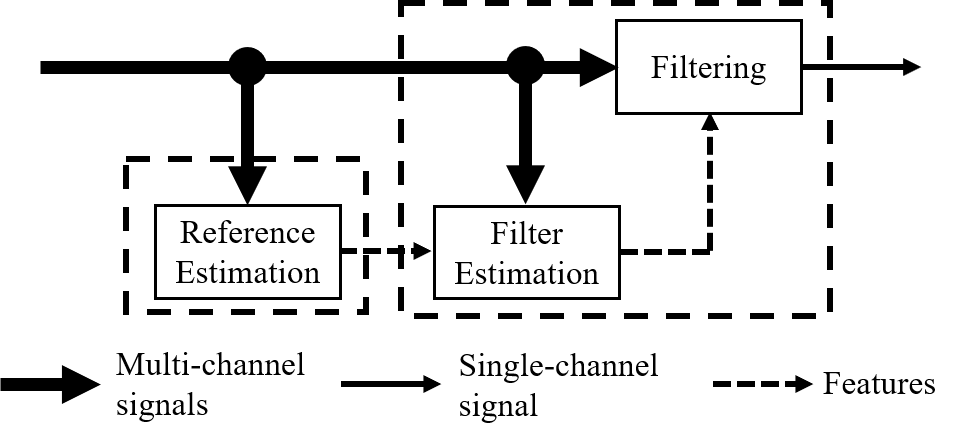}}
  \caption{Basic structure of proposed beamformer}
\label{fig:basic_structure}
\end{figure}
Figure \ref{fig:basic_structure} presents the basic structure of our proposed beamformer. It comprises two DNN blocks; the first estimates a {\em reference} time-frequency representation of the signal, which may be viewed as an amplitude spectrogram, with a time-frequency mask already applied to highlight high-SNR regions.
The second block combines this reference with the multi-channel input signals to compute complex, time-varying beamforming filters.

We use a {\em U-Net} to model each of the two DNN segments -- the choice is predicated in part on the success of these networks in other image and speech processing tasks \cite{ronneberger2015u,jansson2017singing}. 
The U-Net comprises a series of convolutional layers of decreasing size, which attempt to spatially ``summarize'' the input into a bottleneck, followed by a series of deconvolutional layers that reconstruct the target output from the bottleneck. The U-Net is ideally suited to problems such as denoising, where an output of the same spatial extent as the input must be derived from it, while ``squeezing-out'' the noise in the input. E.g. in  \cite{jansson2017singing} it is demonstrated that U-Net is suitable for separating speech and music signals.

\begin{figure*}[htb]
  \centering
  \includegraphics[width=15.0cm]{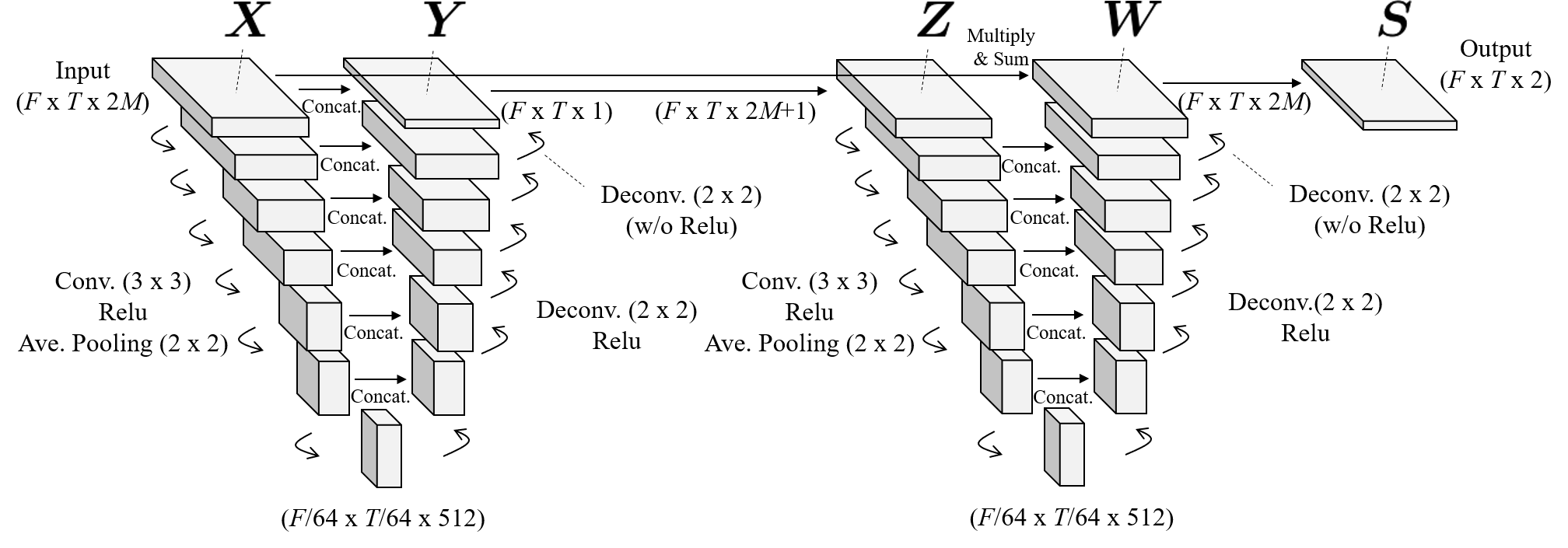}
  \caption{The structure of W-Net Beamformer}
\label{fig:wnet_structure}
\end{figure*}

Fig.\ref{fig:wnet_structure} represents the proposed network structure. The model comprises two U sections; the entire model is called a W-Net \cite{jiang2018w, shi2019w}. We will refer to the model as a W-Net beamformer (W-Net BF). During training each of the U sections is initially separately trained, and finally the entire W is jointly optimized.

In our implementation, in both U segments the convolution layers use kernels of size $3 \times 3$ with a stride of 1, and rectified linear unit (ReLU) activations. Between layers an average pooling operation is performed over blocks of size $2\times 2$ to decrease the spatial span of the input by half in each dimension. A total of 6 convolutional layers are used, with the number of output channels as 16, 32, 64, 128, 256, 512 respectively.
The deconvolution layers use kernels of size $2 \times 2$ with stride 2, and the ReLU activation. Six deconvolution layers are used. The first five layers have output channel counts of 256, 128, 64, 32 and 16 respectively. The last layer has 1 output channel for the first U segment, and $2M$ output channels for the second.

\subsection{Processing array recordings with the W-Net Beamformer}
\label{subsec:whole}
The W-Net beamformer operates on complex time-frequency representations of the speech signal derived using a short-time Fourier transform (STFT). Thus, the input to the beamformer are $X_m(t,f)$, $~t=1,\cdots, T,~f=1,\cdots,F,~m=1,\cdots,M$, the set of STFT coefficients computed from the array recordings.

The magnitude and phase components of $X_m(t,f)$ are concatenated into a $T \times F \times 2M$ tensor $\bm{X}$ as follows:
\begin{equation}
\bm{X}(t,f,m) = |X_m(t,f)|;~\bm{X}(t,f,M+m) = \arg(X_m(t,f)).
\end{equation}
The first U section of the W-Net (which we will refer to as $\text{UNET}_1$) operates on $\bm{X}$ to compute the $T \times F$ matrix $\bm{Y}$.
\begin{equation}
\bm{Y} = \text{UNET}_1(\bm{X}).
\label{eq:unet1}
\end{equation}
$\bm{Y}$ represents an estimated {\em reference} ``clean'' spectral magnitude matrix for the signal. By itself, it has insufficient detail to reconstruct the clean signal; however it is effective as a reference to the second U segment.

The input to the second U segment, which we refer to as $\text{UNET}_2$, is a $T \times F \times (2M+1)$ tensor $\bm{Z}$ formed by concatenating $\bm{X}$ and $\bm{Y}$. The output $\bm{W}$ is a $T \times F \times 2M$ tensor, consisting of the real and imaginary parts of the beamforming filter weights.  
\begin{equation}
\bm{W} = \text{UNET}_2(\bm{Z}).
\label{eq:unet2}
\end{equation}
The filters themselves can be composed from them as $\bm{W}$ as $W_m^{*}(t,f) = \bm{W}(t,f,m) + j\bm{W}(t,f,M+m)$, and $\bm{w}(t,f) = [W_1(t,f), \cdots, W_M(t,f)]^\top$. The final beamformed output is computed as follows:
\begin{equation}
\hat{S}(t,f) = \bm{w}^H(t,f) \bm{x}(t,f).
\end{equation}

\section{Training the W-Net Beamformer}
To train the network, we require collections of ``training'' array recordings, along with an additional {\em noise-free reference} channel (not to be confused with the output of $\text{UNET}_1$), that corresponds to a recording that has been captured by a reference microphone that is influenced by the room impulse response, but not by noise. This is a common requirement for DNN-based beamformer algorithms. This may be recorded, for instance, by a highly directional microphone used only during training, or alternately, for synthetic data, a channel to which noise has not been added. For this reference channel, which we will denote by the subscript $R$, we have $S_{R}(t,f) = H_R(f)S(t,f)$, where $H_R(f)$ is the room impulse response to the reference microphone. 

We train the W-Net beamformer in two stages. In the first we train each of the U blocks independently. Finally we optimize both jointly. 
\subsection{Training $\text{UNET}_1$ independently}
\label{subsec:fp}
To learn the parameters of $\text{UNET}_1$ we minimize the following loss
\begin{equation}
L_1 = \frac{1}{TF}\sum_{t=1}^{T}\sum_{f=1}^F\left(Y(t,f) - |S_{R}(t,f)|\right)^2,
\end{equation}
where $Y(t,f)$ is obtained from $\bm{Y}$.

\subsection{Training $\text{UNET}_2$ independently}
\label{subsec:sp}
To train the parameters of $\text{UNET}_2$, we concatenate $\bm{X}$ and $|S_{R}(t,f)|$ to create $\bm{Z}$, and compute $\hat{S}(t,f)$ from it. Subsequently we minimize the following loss, which uses these values:

\begin{equation}
\begin{split}
L_2 &= \frac{1}{TF}\sum_{t=1}^{T}\sum_{f=1}^F \|\hat{S}(t,f) - S_R(t,f)\|^2.
\end{split}
\label{eq:loss2}
\end{equation}

\subsection{Joint Optimization}
\label{subsec:jt}
Once $\text{UNET}_1$ and $\text{UNET}_2$ are independently estimated, they are finally jointly optimized to minimize loss $L_2$ in Equation \ref{eq:loss2}. Note that backpropagation permits derivatives to propagate from $\text{UNET}_2$ to $\text{UNET}_1$, enabling this optimization.

\subsection{Comparators}
As a comparator (for evaluation), we implemented a direct beamforming filter (without the two-stage structure). This filter too was structured as a U-Net in keeping with our architectures. We refer to it as the U-Net beamformer (U-Net BF), and represent the corresponding network as $\text{UNET}_\text{b}$.
\begin{equation}
\bm{W} = \text{UNET}_\text{b}(\bm{X}).
\label{eq:unet_b}
\end{equation}
The parameters of $\text{UNET}_\text{b}$ were also optimized by minimizing $L_2$. The number of hidden convolutional and deconvolutional layers in $\text{UNET}_\text{b}$ were set to be 1.5 times as many as those for $\text{UNET}_2$, such that the total number of parameters in U-Net BF and W-Net BF were comparable:
the total number of U-Net BF parameters was 5.51 million, while the total number of W-Net BF parameters was 4.90 million.

For comparison, the BLSTM-based mask estimation GEV beamformer\cite{heymann2016neural} (BLSTM-GEV), a state-of-the-art of mask estimation approach, was also evaluated. The GEV beamformer requires a corrective postfilter after beamforming to achieve distortionless response; both the GEV filter and the recommended postfilter were implemented.

\begin{table*}[htb]
\begin{center}
    \caption{Evaluation Result. Results are reported on both test data with both static and moving sources.}
  \label{tab:result}
\scalebox{0.9}{
  \begin{tabular}{|l|c|c|c|c|c|c|c|c|c|c|c|c|c|c|c|}\hline
     \multicolumn{3}{|c|}{Method}  & \multicolumn{8}{|c|}{Evaluation Result} \\ \hline
    & \multicolumn{2}{|c|}{Training Data} & \multicolumn{4}{|c|}{Static-Dataset} &\multicolumn{4}{|c|}{Moving-Dataset} \\ \hline
    & Static & Moving & SNR & SDR & STOI & PESQ & SNR & SDR & STOI & PESQ\\ \hline
    Raw Ch.1 & - & - & 5.51 & 4.92 & 0.90 & 1.66 & 4.96 & 4.91 & 0.87 & 1.49 \\ \hline
    BLSTM-GEV & \checkmark & - & 13.02 & 6.40 & 0.94 & \bf{2.68} & 15.14 & 7.52 & 0.94 & \bf{2.45} \\ \hline
    BLSTM-GEV-m & \checkmark & \checkmark & 12.56 & 5.42 & 0.93 & 2.56 & 14.76 & 6.61 & 0.93 & 2.36 \\ \hline
    U-Net BF & \checkmark & - & 18.70 & 15.89 & \bf{0.96} & 2.57 & - & - & - & - \\ \hline
    W-Net BF $^\dagger$ & \checkmark & - & 16.27 & 14.36 & 0.94 & 2.33 & - & - & - & - \\ \hline
    W-Net BF & \checkmark & - & \bf{18.92} & \bf{16.05} & \bf{0.96} & 2.62 & 16.66 & 14.27 & 0.94 & 2.32 \\ \hline
    W-Net BF-m & \checkmark & \checkmark & 18.63 & 15.96 & \bf{0.96} & 2.61 & \bf{17.00} & \bf{14.46} & \bf{0.95} & 2.37 \\ \hline
  \end{tabular}
 }
  \end{center}
\end{table*}

\end{spacing}
\begin{spacing}{0.91}

\section{EXPERIMENTS}
\label{sec:typestyle}

\subsection{Original Dataset}
In order to test the proposed method in a sufficient variety of conditions, we constructed two sets of simulated microphone array recordings: ``Static-Dataset'' with static noise sources, and a Moving-Dataset with moving sources. (We use synthetic data since standard ``real'' datasets do not contain the variety of conditions we target). 
In both we consider a rectangular room of dimensions $(l_x,l_y,l_z)$, with the origin at one corner.
A circular array with a diameter of 9.26 cm with six microphones ($M=6$) was located at $(l_x/4, l_y/2, 0.5)$.
The impulse responses from a randomly-determined point in the room to the microphone array was generated by the image-source method \cite{lehmann2008prediction, scheibler2018pyroomacoustics}.
For training, we randomly set $l_x,l_y,l_z$ in the range $l_x \in (6.0\text{m}, 9.0\text{m}), l_y \in (4.0\text{m},7.0\text{m}), l_z \in (2.5\text{m}, 3.5\text{m})$, reflection coefficient $\beta$ in the range $\beta \in (0.2,0.8)$, and reflection order $\gamma$ as $17$.
For validation and testing, we set $l_x=8.0\text{m},l_y=6.0\text{m},l_z=3.0\text{m},\beta=0.45,\gamma=17$ and 
$l_x=7.0\text{m},l_y=7.0\text{m},l_z=2.8\text{m},\beta=0.4,\gamma=17$ respectively.
We used the Wall Street Journal (WSJ0) corpus\cite{garofalo2007csr} for the speech source, \textit{si\_tr\_s} for training, \textit{si\_dt\_05} for validation, and \textit{si\_et\_05} for testing.
We also used ESC-50\cite{piczak2015dataset} for the noise source and split it 80\% for training, 10\% for validation, 10\% for testing per sound class.
One speech signal was generated by convolving an impulse response with a speech source, and one to three noise signals were generated by simultaneously convolving impulse responses with noise sources.
In Moving-Dataset, each noise source is moved in the $+y$ axis direction at 0.2m/s.

\subsection{Training and Testing}
For training the W-Net BF, we randomly sampled the impulse response, speech source, and noise source independently and simulated the data obtained by the microphone array.
The SNR was set to follow a normal distribution with $N(\text{5dB},(\text{5dB})^2)$. In our experiment, W-Net BF was first trained using the Static-Dataset. Following this, 
the training procedure was halted after using 15,000,000 utterances in each procedure (\ref{subsec:fp}, \ref{subsec:sp}, \ref{subsec:jt}), 
and the parameters that minimize validation error were chosen for evaluation.
We additionally defined W-Net BF-m, which is finetuned for moving sources using training data of both Static-Dataset and Moving-Dataset. The parameters of W-Net BF-m are initialized with the W-Net BF network trained using Static-Dataset.
Finally, we also trained a variant of W-Net BF without the final joint optimization mentioned in \ref{subsec:jt} (defined as W-Net$^\dagger$). This was evaluated on Static-Dataset to verify the contribution of the joint optimization.

We also trained the comparators U-Net BF and the BLSTM-GEV. The data used to train U-Net BF were doubled to train the comparators to match the effective training data usage for the two. 
We also trained BLSTM-GEV-m, a version of BLSTM-GEV optimized for moving sources, using training data from both Static-Dataset and Moving-Dataset.

For validation and testing, we also independently sampled impulse responses, speech sources, and noise sources. A total of 1024 recordings were generated with mean SNR of 5dB. All recordings were sampled to 16 kHz. We computed a 1024-point STFT with a 75\% overlap and ignored the DC component such that $F = 512$. Although the proposed architecture is fully convolutional, and thus $T$ can be set to arbitrary length, $T$ was set to 256 in this experiment. 

For evaluation the test data were processed by the models, and the performance quantified by a number of metrics:  average value of SNR, 
source to distortion ratio (SDR)\cite{vincent2006performance}, 
short-time objective intelligibility measure (STOI)\cite{taal2011algorithm} 
, and perceptual evaluation of speech quality (PESQ)\cite{rix2001perceptual}.
The SNR was calculated as follows:
\begin{equation}
\text{SNR} = 10\log_{10}\left[\frac{\sum_{t=1}^{T}\sum_{f=1}^{F}\left|\tilde{X}(t,f)\right|^2}{\sum_{t=1}^{T}\sum_{f=1}^{F}\left|\tilde{N}(t,f)\right|^2}\right],
\label{eq:snr}
\end{equation}
where $\tilde{X}(t,f)$ and $\tilde{N}(t,f)$
are STFT coefficients of clean speech and clean noise, respectively, filtered by the beamforming filter.
$S_R(t,f)$ was utilized as a reference signal to calculate SDR, STOI, PESQ.

\subsection{Results}
\label{subsec:results}
The evaluation results are shown in Table \ref{tab:result}. 
The values of SNR and SDR are represented in dB.
It is shown that both the U-Net BF and W-Net BF are superior to the BLSTM-GEV in terms of SNR, SDR, and STOI in Static-Dataset.
In particular, it can be concluded that the DNN-based filter estimation approach can estimate speech signal with less distortion because SDR in U-Net BF and W-Net BF are dramatically improved. 
Upon comparing the W-Net BF$^\dagger$ with W-Net BF in Static-Dataset, it is found that the joint training mentioned in \ref{subsec:jt} is helpful for filling the gap between the two U-Nets, thereby giving rise to good performance.
It can be seen that the W-Net BF consistently outperforms the U-Net BF. 
This indicates that the architecture of the W-Net BF allows for effective computation of the beamforming filter. 
In line with this, we will explore the feasibility of W-Net BF for improving the performance including PESQ in future work.
Upon comparing Static-Dataset with Moving-Dataset, it can be found that the performance in response to moving noise sources of the W-Net BF can be improved by training with a moving noise source, although the performance of BLSTM-GEV cannot be improved, showing that the time-varying beamforming filters from W-Net BF are better at handling spatially non-stationary noise sources.

\section{CONCLUSION}
\label{sec:conclusion}
We have proposed a novel DNN-based beamformer approach called the W-Net Beamformer, that combines the best features of mask-estimation beamformers and filter-estimation beamformers. 
It combines a reference estimation module inspired by the former, and a beamforming filter estimation module inspired by the latter. 
Comparative evaluations showed that our proposed method outperforms both approaches.
In future work,we expect to expand this approach to other speech signal processing tasks and the datasets including real data, and explore the feasibility of optimizing the BF for other objectives, including PESQ. 
We will also explore joint training with specific models for applications such as automatic speech recognition and 
speaker recognition systems. 
 
\end{spacing}

\vfill\pagebreak

\begin{spacing}{0.94}
\bibliographystyle{IEEEbib}
\bibliography{nnbf_short}
\end{spacing}

\end{document}